\documentclass[conference]{IEEEtran}
\IEEEoverridecommandlockouts
\usepackage{Formatting/formatting}
\usepackage{multirow}
\usepackage{colortbl}
\usepackage{booktabs}
\usepackage{tabularx}
\usepackage{caption}
\usepackage{subcaption}

\newcommand{\makeIoULossFigs}{
   \begin{figure*}[ht]
	\centering
	\begin{subfigure}{0.49\linewidth}
		\includegraphics[width=\linewidth]{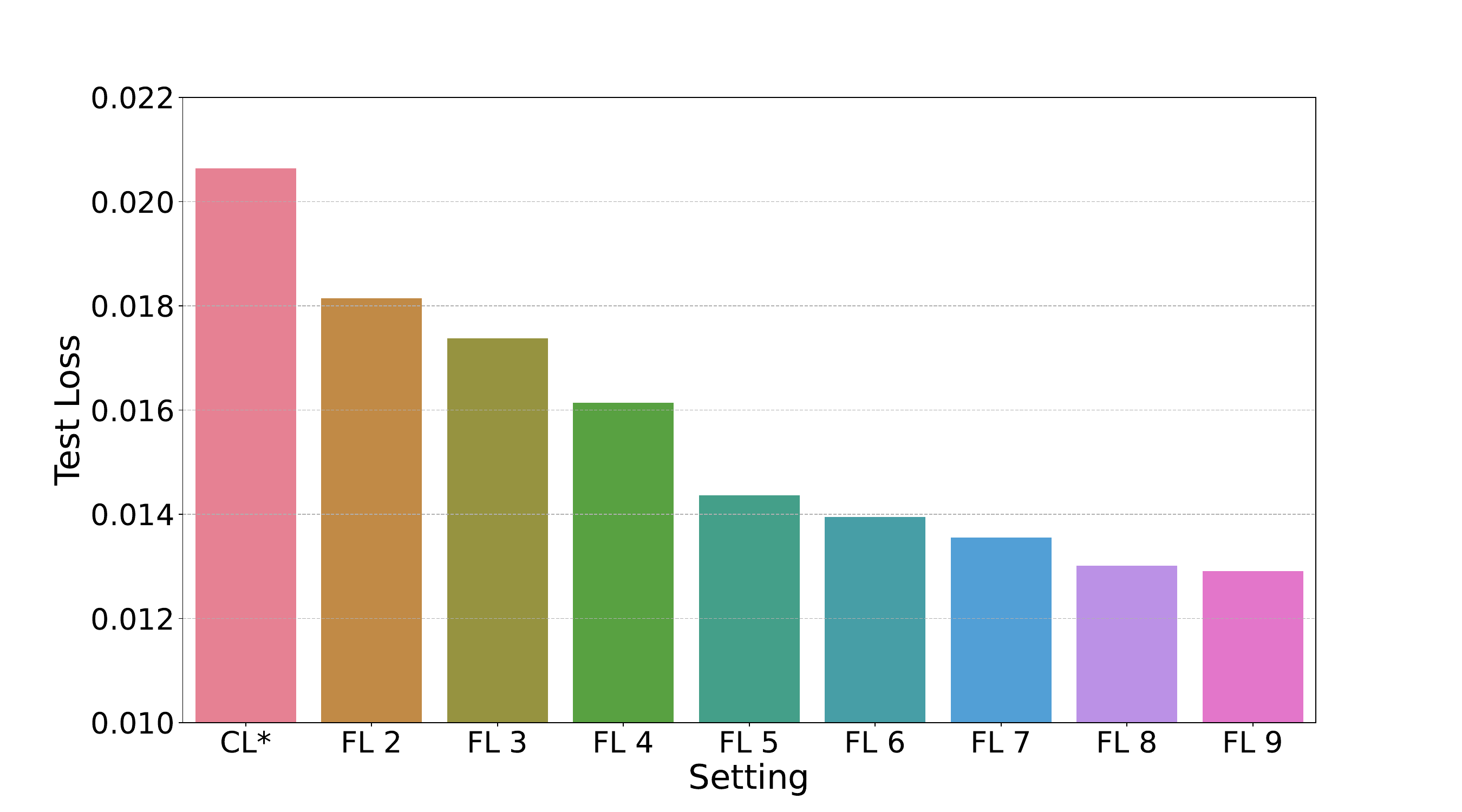}
		\caption{Test Loss}
		\label{fig:subfigA}
	\end{subfigure}
	\begin{subfigure}{0.49\linewidth}
		\includegraphics[width=\linewidth]{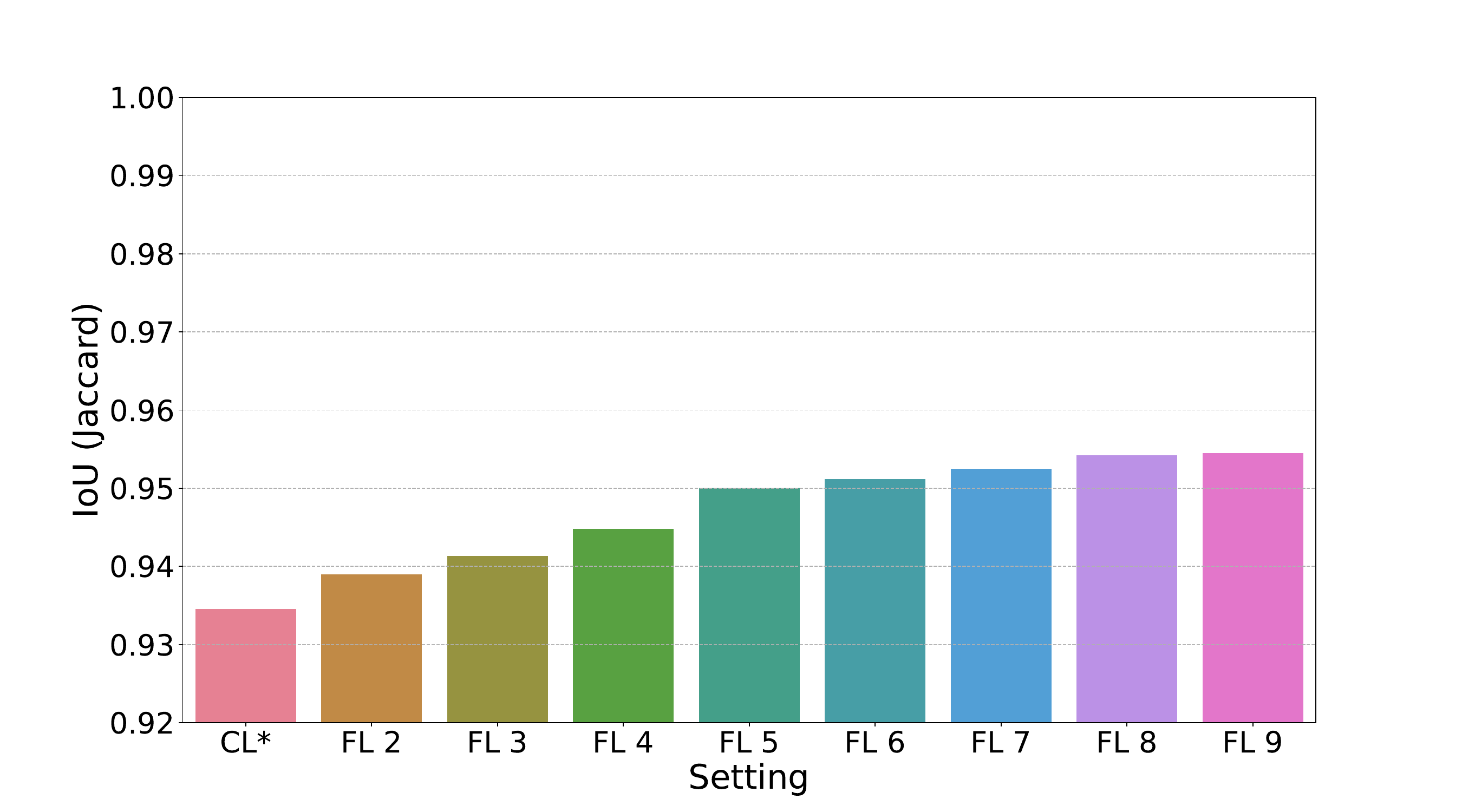}
		\caption{IoU (Jaccard)}
		\label{fig:subfigB}
	\end{subfigure}
	\caption{Test Loss and IoU comparison between Centralized and Federated learning settings using a 32nm technology node dataset. \\ CL$*$: centralized training with 10 samples \\ FL \#: federated training with \# of clients, each with 10 samples 
    \\ Results are reported as mean over 5 independent experiments.}
	\label{fig:results_subfigures}
\end{figure*}}

\newcommand{\makeFLflowFig}{
\begin{figure}[t]
\centering
\includegraphics[width=1\linewidth]{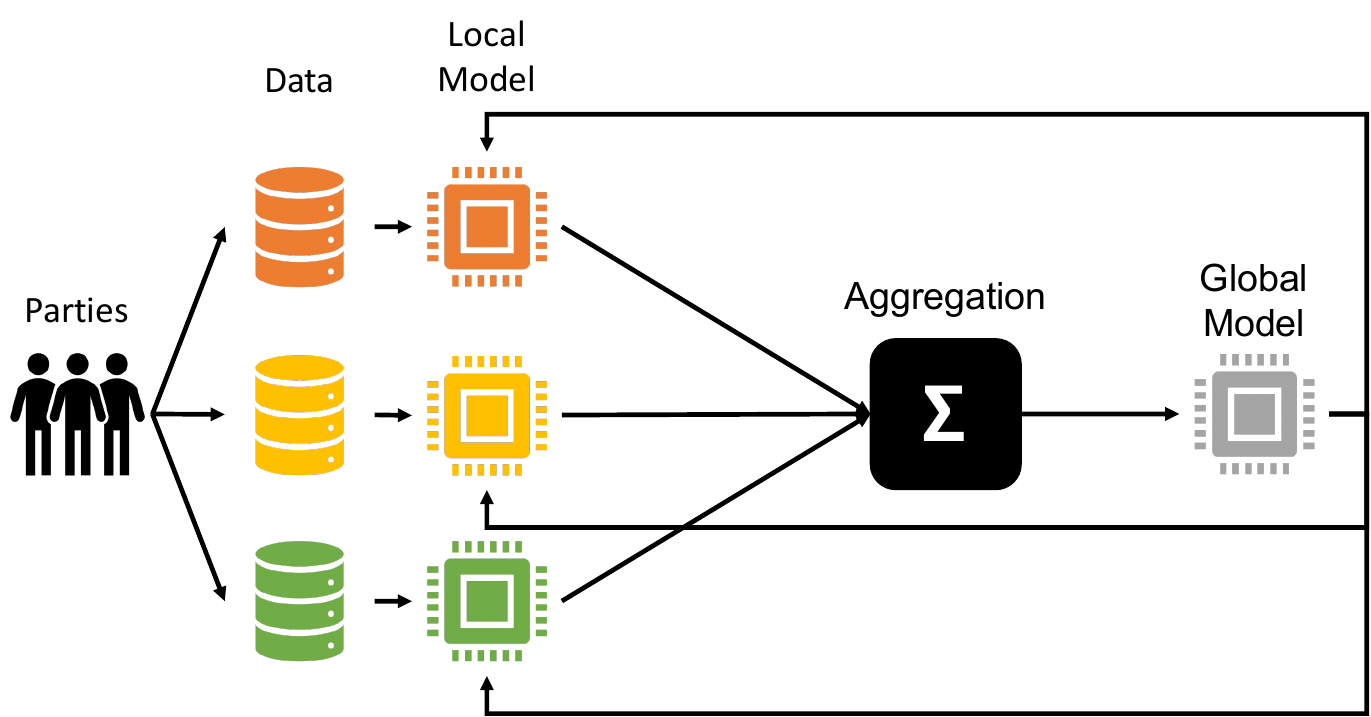}
\caption{Overall workflow of federated learning. \newline
}\label{fig:Federated_learning}
\end{figure}}

\newcommand{\makeExpprocessFig}{
\begin{figure*}[ht]
\centering
\includegraphics[width=1\linewidth]{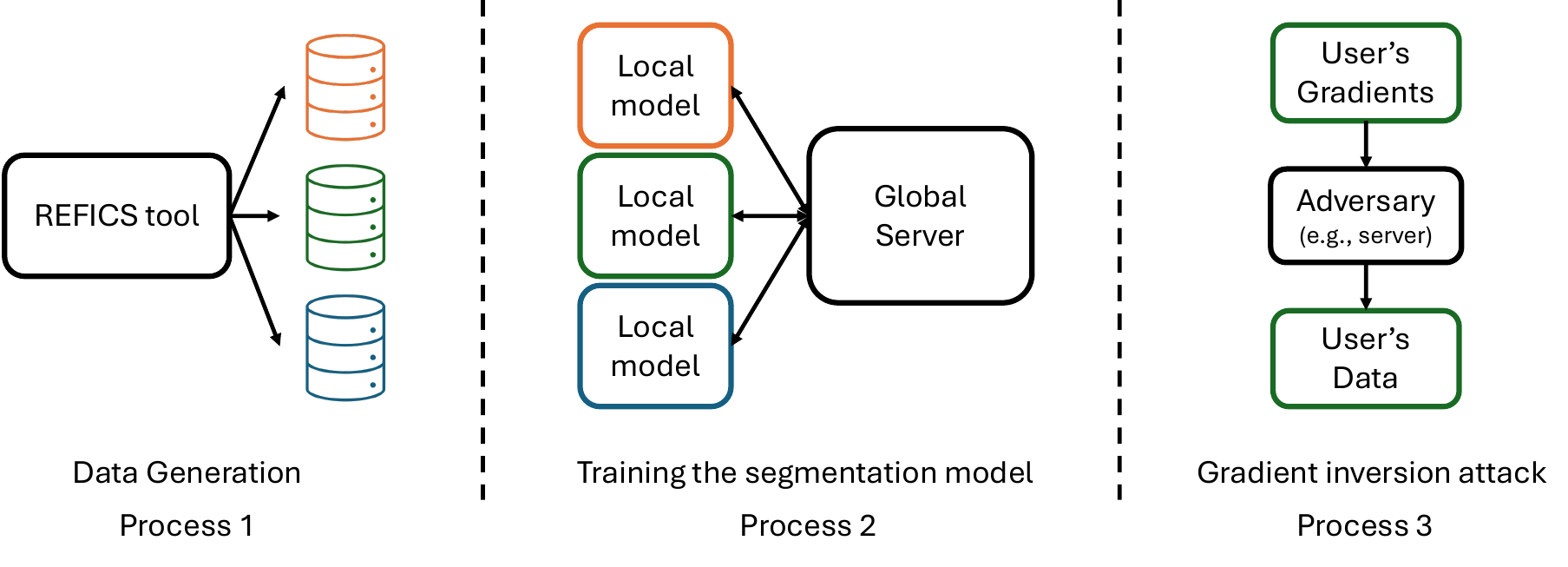}
\caption{Processes for the experiment.}
\label{fig:exp_process}
\end{figure*}}

\newcommand{\makeGradientInversionnewFig}{
\begin{figure*}[ht]
\centering
\includegraphics[width=1\linewidth]{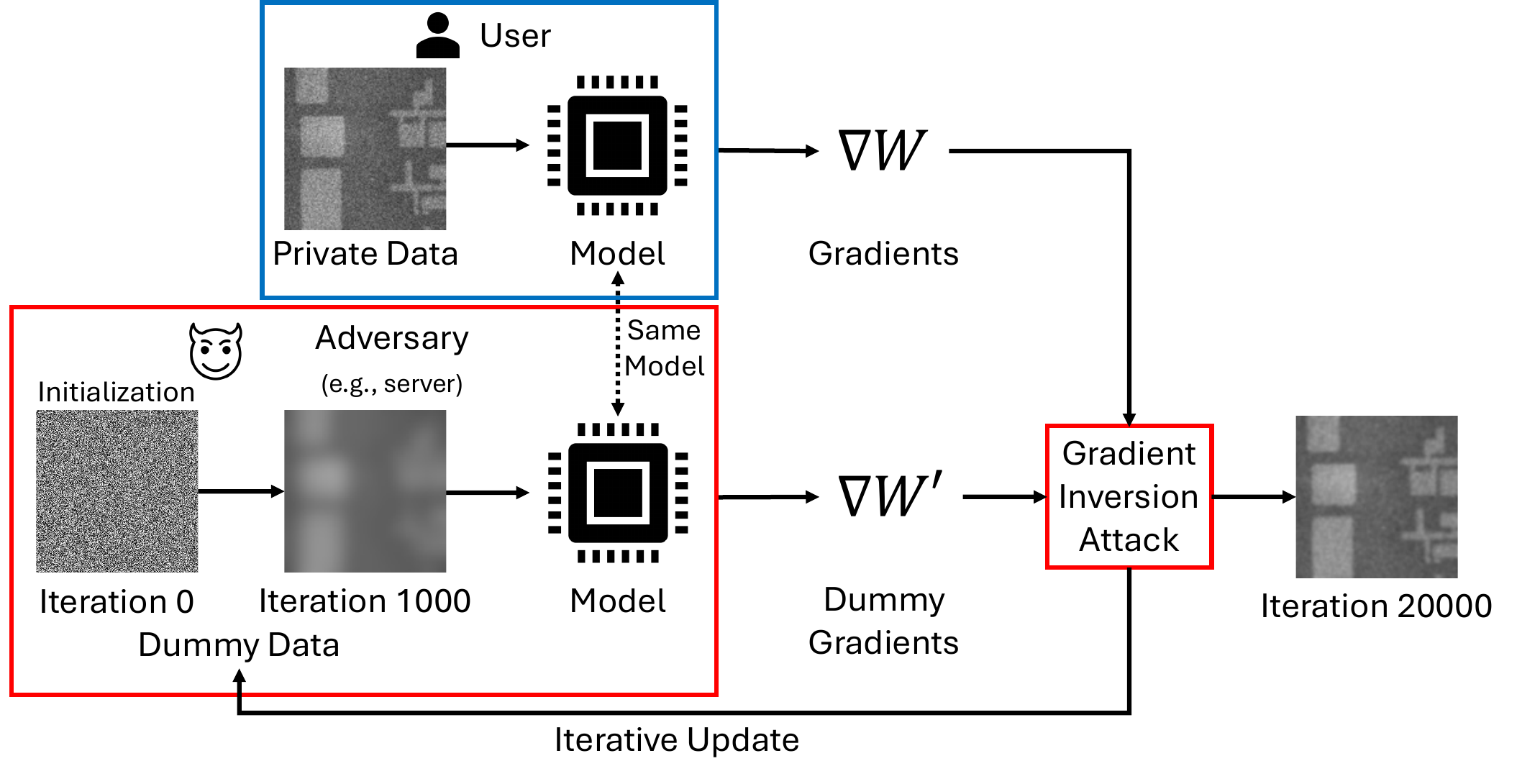}
\caption{Gradient Inversion Attacks}
\label{fig:Gradient Inversion Attacks}
\end{figure*}}

\newcommand{\makeExamplegiaFig}{
\begin{figure}[ht]
\centering
\includegraphics[width=1\linewidth]{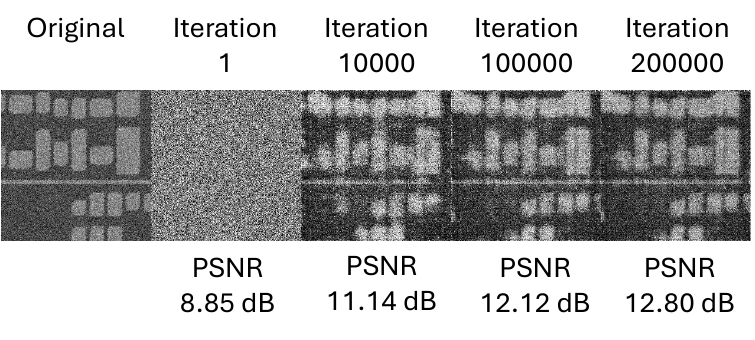}
\caption{Example evolution of the gradient inversion attack on 32nm technology node data. On the left is the original private image, whereas the following images are the reconstructions at different attack iterations.
}\label{fig:example_gia}
\end{figure}}

\newcommand{\makeExampleMSEPSNRFig}{
\begin{figure}[ht]
\centering
\includegraphics[width=1\linewidth]{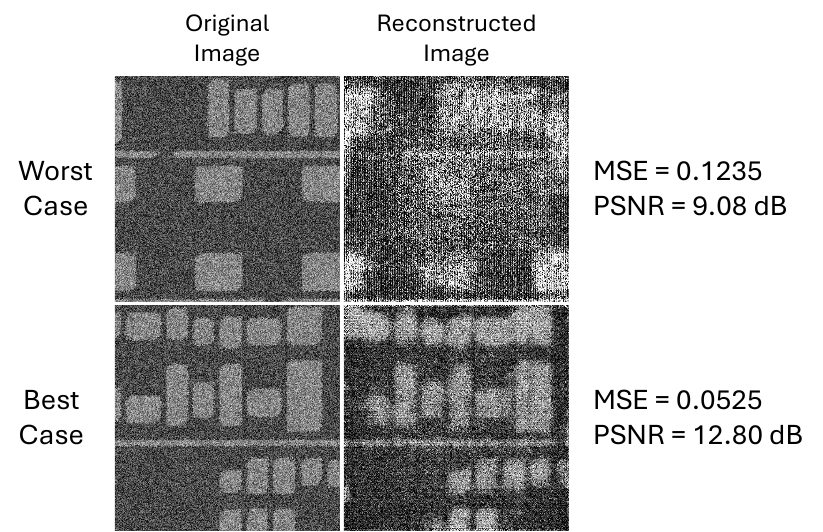}
\caption{Worst and Best reconstructed images for calculating MSE and PSNR.
}\label{fig:example_MSEPNSR}
\end{figure}
}

\newcommand{\makeExampleSSIMFig}{
\begin{figure*}[ht]
\centering
\captionsetup{justification=centering}
\includegraphics[width=1\linewidth]{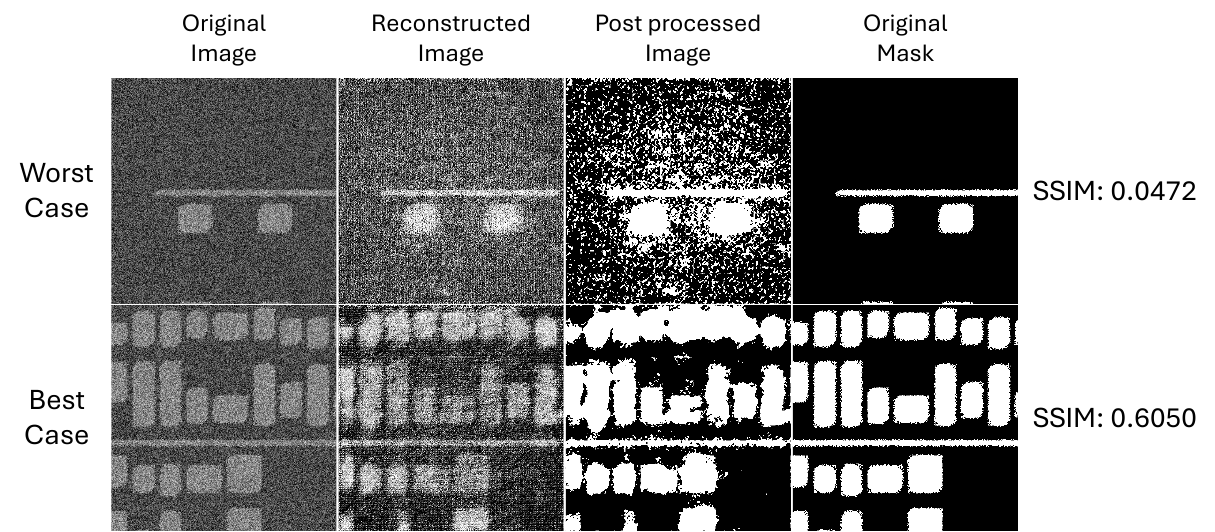}
\caption{Worst and Best reconstructed images for calculating SSIM.\\Post processed images are applied by an algorithm of \cite{wilson2020lasre}. 
}\label{fig:example_SSIM}
\end{figure*}
}

\newcommand{\makeExamplesegFig}{
\begin{figure}[t]
\centering
\includegraphics[width=1\linewidth]{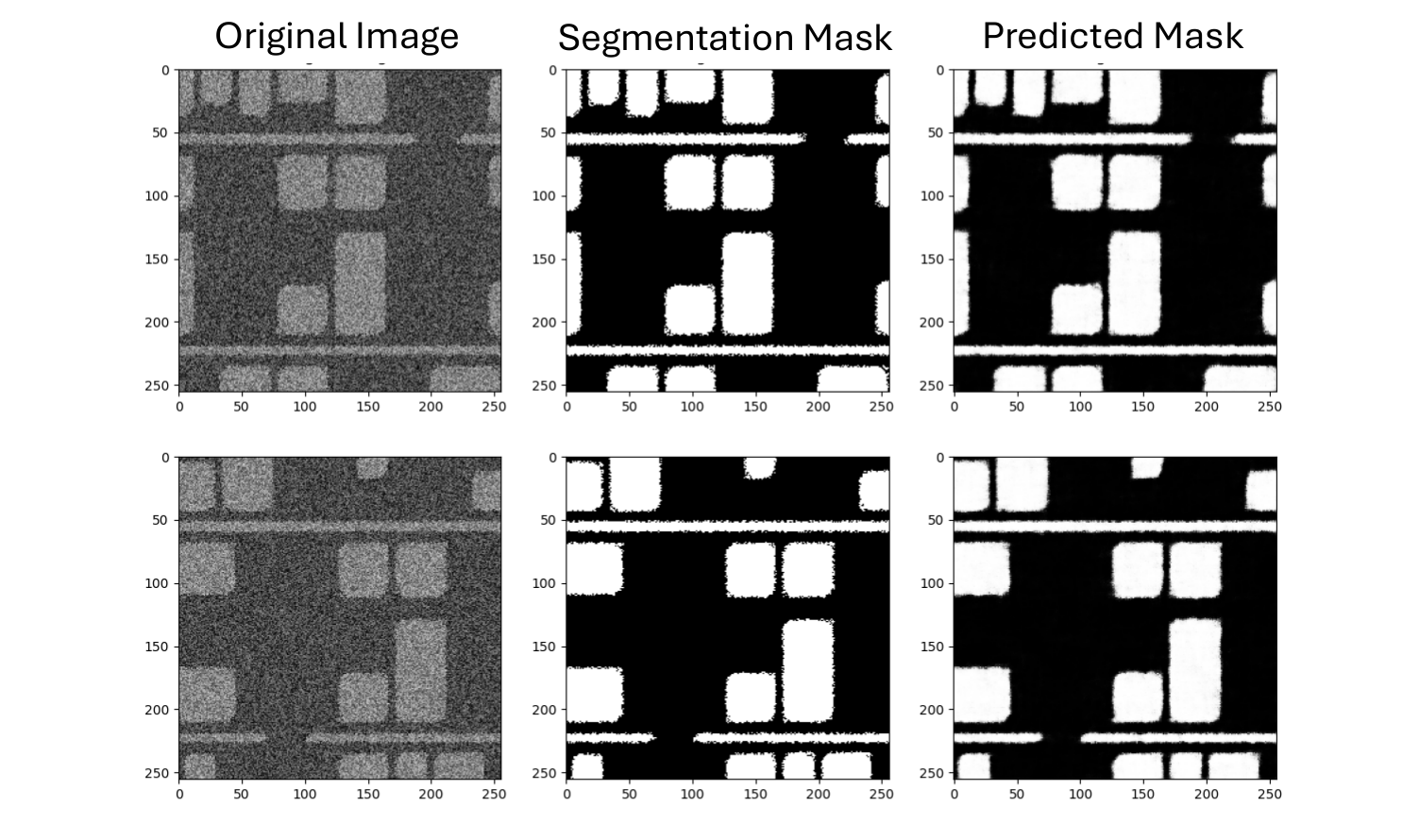}
\caption{Example segmentation result in FL, with 32nm technology node. 
}\label{fig:example_seg}
\end{figure}}

\newcommand{\makeGiametrics}{
\begin{table}[t]
\caption{The average outcome of the gradient inversion attack for 100 samples.}
\centering
\begin{tabular}{ccccc}
\toprule
\textbf{Images} & \textbf{MSE} & \textbf{PSNR} & \textbf{SSIM} \\ 
\midrule
100 images & 0.0830 ± 0.0138 & 10.87 dB ± 0.7387 & 0.2918 ± 0.1114 \\
\bottomrule
\end{tabular}
\label{tab:examplegiametrics}
\end{table}}

\newcommand{\makeExampleresultstable}{

\begin{table}[ht]
\centering
\caption{Metrics for different number of clients in Federated Learning (FL)}
\label{tab:resultsmetrics}
\begin{tabularx}{\textwidth}{c|c|X|X|X}
Number of client's data & Total number of clients in FL & IoU & Dice-Coefficient & MSE \\
\hline
200 & 1 & 0.37121 & 0.29220 & 0.16925 \\
     & 2 & 0.37121 & 0.26779 & 0.17907 \\
     & 3 & 0.37121 & 0.31607 & 0.15716 \\
     & 4 & 0.37121 & 0.31363 & 0.15738 \\
\hline
500 & 1 & 0.99187 & 0.91161 & 0.00714 \\
     & 2 & 0.99053 & 0.89875 & 0.00827 \\
     & 3 & 0.99690 & 0.92868 & 0.00415 \\
     & 4 & 0.99793 & 0.94672 & 0.00278 \\
\hline
800 & 1 & 0.99961 & 0.99298 & 0.00022 \\
     & 2 & 0.99967 & 0.99420 & 0.00016 \\
     & 3 & 0.99968 & 0.99501 & 0.00016 \\
     & 4 & 0.99967 & 0.99367 & 0.00019 \\
\end{tabular}
\end{table}
}

\newcommand{\makeSegresultss}{\begin{table*}[ht]
\newrobustcmd{\B}{\bfseries}
    \caption{Performance comparison between Centralized and Federated learning settings using a 32nm technology node dataset. \\ Metrics are reported as mean $\pm$ standard deviation over 5 independent experiments.}
    \centering
    \begin{tabular}{cccccc}
        \toprule
        \textbf{Setting} & \textbf{Total Number of Data} & \multicolumn{4}{c}{\textbf{Metrics (Mean ± SD)}} \\
        \cmidrule(lr){3-6}
        & & \textbf{MSE} & \textbf{IoU (Jaccard)} & \textbf{SSIM} & \textbf{Total training time (sec)} \\
        \midrule
        CL 1 client & 10 & \B 0.02064 ± 0.00151 & \B 0.93452 ± 0.00419 & \B 0.69728 ± 0.00809 & \B 52.78 ± 19.73 \\
        FL 2 clients & 20 & 0.01815 ± 0.00159 & 0.93895 ± 0.00459 & 0.83387 ± 0.01033 & 175.90 ± 16.27 \\
        FL 3 clients & 30 & 0.01738 ± 0.00029 & 0.94134 ± 0.00084 & 0.83873 ± 0.00641 & 186.80 ± 21.66 \\
        FL 4 clients & 40 & 0.01614 ± 0.00048 & 0.94476 ± 0.00144 & 0.83580 ± 0.01229 & 188.82 ± 27.28 \\
        FL 5 clients & 50 & 0.01437 ± 0.00052 & 0.95006 ± 0.00154 & 0.85933 ± 0.01275 & 199.50 ± 33.15 \\
        FL 6 clients & 60 & 0.01395 ± 0.00053 & 0.95119 ± 0.00161 & 0.85793 ± 0.00812 & 215.51 ± 30.33 \\
        FL 7 clients & 70 & 0.01355 ± 0.00009 & 0.95247 ± 0.00033 & 0.85591 ± 0.00546 & 221.54 ± 38.89 \\
        FL 8 clients & 80 & 0.01301 ± 0.00037 & 0.95425 ± 0.00115 & 0.85288 ± 0.00348 & 219.94 ± 23.71 \\
        FL 9 clients & 90 & \B 0.01291 ± 0.00020 & \B 0.95452 ± 0.00065 & \B 0.85858 ± 0.00626 & \B 221.78 ± 22.13 \\
        CL 1 client & 90 & \B 0.01312 ± 0.00046 & \B 0.95657 ± 0.00141 & \B 0.90564 ± 0.00574 & \B 217.99 ± 27.65 \\
        \bottomrule
    \end{tabular}
    \label{tab:FLsegresult_combined}
\end{table*}}

\newcommand{\makeHyperparameter}{
\begin{table}[t]
\caption{Hyperparameter settings for Centralized Learning (CL) and Federated Learning (FL).}
\centering
\begin{tabular}{ccc}
\toprule
Hyperparameter & CL & FL \\
\midrule
Total Epochs & 100 & 100 \\
\midrule
Local Epochs & - & 1 \\
\midrule
Server Rounds & - & 100 \\
\midrule
Learning Rate & 0.00008 & 0.00008 \\
\midrule
Batch Size & 1 & 1 \\
\bottomrule
\end{tabular}
\label{tab:hyperparameters}
\end{table}
}

\def\BibTeX{{\rm B\kern-.05em{\sc i\kern-.025em b}\kern-.08em
    T\kern-.1667em\lower.7ex\hbox{E}\kern-.125emX}}
\begin{document}


\title{Potentials and Pitfalls of Applying Federated Learning in Hardware Assurance}
\author{
\IEEEauthorblockN{Gijung Lee, Wavid Bowman, Olivia Dizon-Paradis, Reiner Dizon-Paradis, \\Ronald Wilson, Damon Woodard, Domenic Forte}

\IEEEauthorblockA{Department of Electrical and Computer Engineering, University of Florida, Gainesville, FL, USA\\
Email: \{lee.gijung, wavid.bowman, paradiso, reinerdizon, ronaldwilson\}@ufl.edu, \{dwoodard, dforte\}@ece.ufl.edu\\}}

\maketitle

\begin{abstract}
As microelectronics flourish and outsourcing of the design and manufacturing stages of integrated circuits (ICs) and printed circuit boards (PCBs) becomes the norm, microelectronics stakeholders must also confront a new wave of security challenges, including the threats posed by hardware Trojans, counterfeit electronics, and reverse engineering attacks. Traditional detection and prevention methods like testing and side-channel analysis have limitations in reliability and scalability. Automated reverse engineering by deep learning (DL) models is a foolproof approach to hardware assurance, but faces challenges due to limited data. By pooling data from different stakeholders (competitors in industry, governments, etc.), DL models can be more effectively trained but privacy of intellectual property (IP) is a significant concern. Federated Learning (FL) has been proposed as a potential alternative allowing for the collaborative training of a DL model without sharing raw data. While FL has been widely used in healthcare, IoT, and finance, its application in hardware assurance remains underexplored. This study investigates, for the first time, FL-based DL for hardware assurance, demonstrating that FL outperforms single-client centralized learning in segmentation tasks for reverse engineering. Our results show that increasing the number of clients improves FL performance by collaboratively training the model with more data. However, and more importantly, a major pitfall of FL is also exposed -- it remains vulnerable to gradient inversion attacks. We show that SEM images used in FL can be recovered by attackers, which would therefore expose the sensitive and proprietary IPs that FL was supposed to protect. We highlight these privacy risks and also suggest future research directions to improve security and effectiveness in hardware assurance.

\end{abstract}


\section{Introduction}\label{sec:introduction}
While microelectronics technology has rapidly proliferated in modern life due to reduced development time and costs, it has also increased security threats~\cite{botero2021hardware}, such as hardware Trojans and the rise of counterfeit electronic components.  Untrusted foundries may overproduce ICs \cite{tehranipoor2015counterfeit} or even leak/steal IP. Attackers can employ reverse engineering (RE) techniques to recover semiconductor intellectual property (IP); this allows them to build/source counterfeit integrated circuits (ICs) that are unsafe, which can be costly to IP owners and reduce consumer confidence. Hardware assurance approaches aim to verify device integrity and authenticity, but existing techniques face challenges, such as increased resource demands, vulnerability to tampering, and low reliability in detecting small-scale Trojans.
Hardware assurance approaches help ensure some level of trust by showing that the device has not been maliciously modified and confirming that the device is not counterfeit. However, existing techniques are limited and ineffective in achieving assurance and trust.

As data-driven paradigms (e.g., machine learning and deep learning) are increasingly used in hardware assurance, particularly verification by \textit{automated reverse engineering (RE)}, a number of vulnerabilities arise due to the incorporation of sensitive IP in these workflows. IC designers, entities, and third parties are reluctant to share their sensitive or private data with others since it could in turn be stolen or analyzed to identify vulnerabilities. This motivates the need for other data-driven paradigms that address data privacy concerns. One such paradigm, Federated Learning (FL), is seen as an innovative approach to address these issues with the use of distributed learning across multiple participants. Each participant keeps their data locally and aggregated models are created on a central server. This approach aims to maintain a balance between collaboration and model performance while protecting privacy. FL has been actively researched and utilized in a variety of fields, such as healthcare, IoT, and finance, but it has received relatively little attention in hardware assurance, and there are few practical applications in the field. 
In this study, we examine a use case for FL in hardware assurance.

FL is often regarded as a privacy-preserving technique, but recent research threatens this assumption. Notably, Gradient Inversion Attacks (GIA) underscore the privacy limitations of FL and demonstrate that the sensitivity of data and model updates can be exploited.
Attacks like these are a major factor in weakening FL's practical privacy proposition.
This study focuses on situations where all participants use the deployed model without alterations to the data or the model itself (i.e., ``honest-but-curious'' setting \cite{goldreich2001foundations, paverd2014modelling}). We aim to verify the benefits of FL in hardware assurance while examining the feasibility of gradient inversion attacks that may counter those benefits.
The paper makes the following contributions:
\begin{itemize}
    \item To the best of our knowledge, we are the first to apply Federated Learning (FL) in hardware assurance using deep learning for a segmentation task on Scanning Electron Microscope (SEM) images.
    \item We examine the advantages and challenges of applying FL to hardware assurance, highlighting it alone cannot prevent privacy threats. Key considerations of privacy threat (e.g., GIA) include combining the two loss functions, the approximation of gradients derived from model weights, use of a more complex model and a high-resolution image that made success of the GIA harder, and the realistic attack scenario where the attacker has limited access to a ground truth mask.
    \item We offer suggestions for future research directions to explore other privacy-preserving techniques and attacks on AI systems.
\end{itemize}

The remainder of this paper is organized as follows. Section \ref{sec:background} provides background on (1) hardware assurance with deep learning; (2) federated learning and its benefits and limitations for hardware assurance; and (3) segmentation task for IC reverse engineering. The threat model and privacy risks in federated learning are discussed in Section~\ref{sec:threat}.
Section \ref{sec:methodology} describes the methodology, including segmentation tasks in federated learning and the gradient inversion attack. Section \ref{sec:results} presents the experimental results. Section \ref{sec:Discussion and Future Research Directions} discusses potential future research directions. Finally, Section \ref{sec:conclusions} summarizes the study’s conclusions.

\section{Background}\label{sec:background}
This section provides overviews of hardware assurance using deep learning and its challenges, as well as federated learning and its benefits and limitations in hardware assurance.

\subsection{Deep Learning-based Hardware Assurance}\label{Hardware assurance using deep learning:background}

Due to increasing demands and the growing complexity of ICs, many semiconductor manufacturers rely on third-party IP and external vendors for multiple phases of the IC supply chain \cite{sharma2021new}. This collaboration has enhanced efficiency and reduced costs; however, it has also introduced significant security risks, including hardware Trojans, IP piracy, IC counterfeiting, and IC overproduction \cite{rostami2014primer}. Ensuring that electronic hardware functions as intended and remains free from vulnerabilities or malicious modifications is critical, making hardware assurance a key area of research. In fact, the CHIPS Act~\cite{senate} reinforces this by promoting domestic semiconductor manufacturing and secure supply chains. It also encourages investment in security technologies and fosters collaboration among industry, academia, and government to develop best practices and standards for ensuring hardware integrity and reliability.

Deep learning (DL)-based hardware assurance methods address limitations in traditional techniques by enhancing detection capabilities and automating security analysis. One of the critical areas where DL has been applied is hardware Trojan detection. Hardware Trojans, which are malicious modifications inserted at multiple phases of the IC supply chain, compromise chip security and functionality. For example, \cite{sankaran2021deep} developed DL models using neural networks and an autoencoder to detect Trojans that existing approaches, such as support vector machines (SVMs) and K-nearest neighbors (KNN) classifiers, could not detect. \cite{sharma2021new} proposed the deep convolutional neural networks (CNNs) to detect Trojans from IC layout images.

Beyond hardware Trojan detection, DL has been widely adopted in other hardware assurance applications, including fault detection and IC reverse engineering (RE). Fault detection is essential in industries like aerospace, automotive diagnostics, and industrial automation, where undetected defects can have severe consequences. Traditionally, fault detection has relied on experienced labor, which is time consuming and prone to human error \cite{jha2023deep}. \cite{shu2021quality} addressed these challenges by introducing a parallel deep convolutional model, the Parallel Spatial Pyramid Pooling Network (PSPP-net), for LED chip defect classification, achieving high accuracy in defect identification. Meanwhile, RE is another crucial aspect of hardware assurance, used to detect hardware Trojans and verify IP compliance through destructive physical analysis. Recent research has leveraged DL models for IC reverse engineering, particularly in analyzing scanning electron microscopy (SEM) images. \cite{lin2021deep} utilized Fully Convolutional Networks (FCNs) with a VGG-16 backbone to segment vias, contacts, and metal lines in SEM images, reducing annotation errors compared to traditional methods. \cite{lin2023sem2gds} introduced the SEM2GDS framework, a two-stage deep learning pipeline that converts SEM images into GDS layouts using U-Net-based segmentation models. Image-to-image translation using generative adversarial networks (pix2pix and cycleGAN) and blind denoising models (DnCNN and CBDNet) were also explored for segmentation. Although both types of DL methods produced near perfect results based on metrics such as SSIM, IoU, CC-US/OS, denoising models produced better overall segmentation results due to their understanding of residual noise in image \cite{wilson2021refics, wilson2022refics}. 

Machine learning, and in particular DL models require substantial amounts of data to effectively train. Given the extensive time and resources required to de-layer and perform SEM images on a single chip, let alone many chips in each technology node, obtaining this data and thus generating effective models is a challenge for any single organization.

\makeFLflowFig

\subsection{Federated Learning}\label{Federated Learning:background}

Federated Learning (FL) \cite{Wen2023federated} is a collaborative machine learning approach where parties work together without sharing private data. Instead of training on a centralized server, each party locally trains the model and shares only model updates. This process protects data privacy by preventing potential data leaks stemming from sharing private data for training.  FL also enables more secure collaboration among organizations, such as IP vendors, manufacturers, and researchers, allowing them to develop DL models without (ideally) exposing confidential data or IP. As shown in Fig. \ref{fig:Federated_learning}, FL follows these main steps \cite{torkzadehmahani2022privacy}:
\begin{enumerate}
    \item For the current iteration of the model (\textit{global model}), the server chooses a group of parties to participate.
    \item The server provides the global model, which is the most recent version, to the chosen parties.
    \item Every chosen party calculates the local parameters with its own private data and the current model (local model), e.g., obtains the \textit{local gradient updates} by running the gradient descent algorithm on its own data, initialized by the current model.
    \item To update the global model, the server gathers local parameters or local gradient updates from the chosen parties and \textit{aggregates} them.
\end{enumerate}

\subsection{Application of FL in DL-based Hardware Assurance}
This section explores potential strengths and challenges of applying FL in hardware assurance using deep learning DL and the limited application of FL in hardware assurance using DL by comparing its use in other domains like healthcare, IoT, and finance.

\subsubsection{Applications in Other Domains}
FL has been successfully implemented across different fields to boost collaboration, improve model performance, and address data privacy concerns. In healthcare, FL helps train DL models for tasks like disease diagnosis and classification. For example, \cite{li2022federated} used FL for breast cancer classification, while \cite{islam2023effectiveness} applied it to brain tumor detection with MRI images. In IoT, FL enables privacy-preserving AI applications in smart cities, smart industries, smart healthcare, and smart transportation by reducing reliance on centralized data processing, which can lead to security risks \cite{nguyen2021federated}. \cite{zhao2019multi} proposed a multi-task deep neural network for a variety of network-related tasks. It performs traffic classification, network anomaly detection, and VPN traffic recognition simultaneously. \cite{etiabi2024femloc} proposed a federated meta-learning framework for localization (FeMLoc) that allows indoor localization systems to achieve higher accuracy and robustness. In the financial industry, machine learning algorithmic models are widely used in various fields such as fraud detection, investment advising, algorithmic trading, robo-advising, and loan screening. However, due to a lack of sufficient data, unreliable alternative data are often used, which can lead to bias in the measurement process. Due to this problem, many researchers in the finance industry focus on a novel approach using FL to address this challenge. \cite{awosika2024transparency} proposed a new approach for financial fraud detection using FL and Explainable AI (XAI). \cite{lee2023federated} presented a FL prototype that allows smaller financial institutions (FIs) to compete with large FIs that have better-performing models for credit risk assessment by training with larger datasets.

\subsubsection{Benefits of FL in DL-based Hardware Assurance}
The successful application of FL and DL in these other fields strongly suggests that they could also have great potential in hardware assurance. 
Similar to the rest of these fields, hardware assurance requires large-scale data and likewise shares the common challenge of simultaneously maintaining data privacy and security. Therefore, improving the hardware assurance process through FL-based DL technology is expected to enhance the reliability of the assurance system and significantly improve the level of cooperation required in the global supply chain. Hardware assurance is essential for evaluating the quality, reliability, and security of electronic devices and semiconductors, and FL can strengthen these in the following ways: 
\begin{itemize}
    \item \textit{Quality and Defect Detection}: DL models can be utilized in image analysis and pattern recognition for hardware defect detection and quality assessment. FL can contribute to improving defect detection performance by learning data from various competing manufacturers (TSMC, Intel, Samsung, etc.).
    \item \textit{Privacy Preservation in Data Integration}: Allows training locally without exposing data, enabling global model learning while protecting sensitive hardware IP, process, and manufacturing data.
    \item \textit{Model Generalization}: By applying FL, models trained on data collected from a variety of manufacturing environments facilitate generalization in hardware assurance processes, providing robust models that can operate under various conditions and technology nodes.
\end{itemize}

\subsubsection{Limitations of FL in DL-based Hardware Assurance}
We investigated the case of applying DL and FL to hardware assurance. For this purpose, related papers and case studies in this field were systematically searched, but few studies using both FL and DL were found. 
Challenges and limitations of applying FL and DL techniques in hardware assurance include:
\begin{itemize}
    \item \textit{Lack of data standardization between stakeholders}: The hardware industry has a complex ecosystem in which multiple stakeholders, including various manufacturers, design companies, and test companies, collaborate. However, the formats and storage methods of data used by each stakeholder are not standardized, making data integration for FL model training difficult.
    \item \textit{Hardware datasets are highly specialized and non-overlapping across stakeholders}: Not only is data in the hardware field highly specialized, but the data held by each stakeholder is unique, with little overlap. The non-overlapping quality of the data makes useful generalizations difficult in FL-based DL model training.
    \item \textit{IP is often the core business value in the hardware field}: In the hardware industry, IP is considered a core business value for companies, and security requirements to protect it are very high. FL offers the advantage of learning models without sharing data directly, but companies have a strong tendency to avoid even the risk of indirect exposure\footnote{Our analysis of gradient inversion attacks later in the paper will substantiate this concern.}. In particular, if sensitive design or data is exposed, there is a possibility of leaking important business information to competitors or external adversaries, making the adoption of FL more difficult.
    \item \textit{Reverse engineering risks in hardware are higher than in healthcare, IoT, and finance}: Hardware RE carries national security concerns absent in the other domains. If a shared model falls into the hands of a malicious user, there is a possibility that the model can be reverse-engineered to extract sensitive data or IP. This increases the risk of leaking technical details of the hardware design and key elements of the assurance system, and makes it difficult to ensure the safety of collaboration through the FL model.
    \item \textit{Use of various technology nodes}: Hardware design and manufacturing processes use different technology nodes (e.g., 32nm, 90nm, etc.). Each technology node has its own data characteristics and production process, and it is very difficult to learn them comprehensively in one FL model. These differences between processes further deepen the heterogeneity of the data and create additional technical challenges during model training.
\end{itemize}

The challenges mentioned above are the main reasons why the application of FL and deep learning in the field of hardware assurance is progressing more slowly compared to other domains. To solve these problems, it is essential to standardize data, develop technologies for IP protection, improve the security of the FL model, and develop an integrated model that encompasses various node technologies. As a first step toward filling this gap, this study seeks to discuss the applicability, potential benefits, and pitfalls of FL and DL in hardware assurance.

\subsection{IC Reverse Engineering with SEM in FL}

IC RE plays a crucial role in hardware assurance by reconstructing circuit layouts from images of physical structures. This process is essential for verifying designs, detecting unauthorized modifications, and ensuring the integrity of semiconductor devices. This process relies heavily on image segmentation, particularly when working with Scanning Electron Microscope (SEM) images to separate key circuit components. Image segmentation is the process of partitioning an image into meaningful areas of interest, facilitating easier analysis and interpretation of its contents by identifying and isolating objects or areas of interest. In the case of hardware RE, segments would include diffusion regions, polysilicon or metal gates, contacts and vias, and metal interconnects. Accurate segmentation remains a challenge due to variations in image quality, noise, and complex semiconductor structures. While various segmentation techniques exist, DL-based approaches such as U-Net have shown superior performance in extracting complex semiconductor structures compared to traditional image processing methods \cite{kalber2021u, wilson2021refics, qiao2025ra}.

Despite the effectiveness of deep learning in image segmentation, applying it to IC reverse engineering raises concerns about data privacy. Semiconductor manufacturers and research institutions often work with proprietary datasets that cannot be shared due to confidentiality constraints. This challenge motivates the exploration of FL, which enables collaborative model training without exposing raw SEM images. By leveraging FL, multiple companies can contribute to improving a segmentation model while preserving the security of their data. 

To the best of our knowledge, no prior work has investigated the use of FL for IC segmentation tasks using SEM images. This gap presents an opportunity to explore how FL can enhance segmentation accuracy while addressing data privacy concerns in semiconductor analysis. In this study, we apply FL to train a U-Net-based segmentation model, aiming to extract functional circuit elements from SEM images. Through this approach, we aim to demonstrate that FL can improve the performance of segmentation models but fails to provide the desired level of data privacy on its own upon deeper inspection.

\begin{table*}[ht]
    \centering
    \renewcommand{\arraystretch}{1.5} 
    \caption{Comparison of Membership Inference Attack (MIA) and Gradient Inversion Attack (GIA).}
    \begin{tabular}
    {@{}>{\centering\arraybackslash}p{1.2cm}>{\raggedright\arraybackslash}p{7.5cm}>{\raggedright\arraybackslash}p{7.5cm}@{}}
        \toprule
        \textbf{Feature} & \textbf{Membership Inference Attack (MIA)} & \textbf{Gradient Inversion Attack (GIA)} \\ \midrule
        \textbf{Goal} & Determine whether a specific data sample was used in training. & Reconstruct input data from shared gradients. \\ 
        \textbf{Attack Method} & Uses model outputs (e.g., confidence scores, loss values) to infer membership. & Exploits gradients in FL or training to reconstruct the original input. \\ 
        \textbf{Target} & Trained model (black-box or white-box). & Shared gradients in distributed learning (e.g., FL). \\ 
        \textbf{Threat Model} & Adversary queries the model with known/unknown data. & Adversary intercepts gradients shared during training. \\ 
        \textbf{Impact} & Privacy leakage about training data presence. & Full reconstruction of private data, causing severe privacy risks. \\ 
        \textbf{Common Defenses} & Differential privacy, adversarial regularization, confidence masking. & Gradient clipping, differential privacy, secure aggregation. \\ 
       \textbf{Use Case} & Checking if PDK or technology node was used to train a model & Extracting a SEM image from a federated learning update. \\ 
        \textbf{Severity} & Moderate (leaks membership info). & High (can recover full private data). \\ \bottomrule
    \end{tabular}
    \label{tab:mia_vs_gia}
\end{table*}

\makeGradientInversionnewFig


\section{Threat Model} \label{sec:threat}
FL models are deployed in a distributive configuration, dispersing their risks across multiple devices. Without deference to which device gets affected with potential threats in FL systems, both the server (e.g., central) or client (e.g., user) models are considered untrusted entities. In addition to the devices and models themselves, attackers can exploit the communication among these devices via bus snooping or man-in-the-middle attacks to gather model or data information, such as weights, biases, and gradients/losses. This is especially effective when communication is not encrypted. There are also multiple vulnerabilities throughout the machine learning pipeline. During the model training phase, model or data information can be stolen during transmission between the different devices in an FL system. In addition, during the model deployment phase (i.e., post-model training), gradients/losses can be leaked during transmission and used to reconstruct sensitive information about the dataset or the model.

\subsection{Privacy Threats in FL}
FL is attracting attention as an innovative technology that can learn high-performance models while ensuring data privacy. However, FL alone cannot completely prevent all types of privacy violations and may be vulnerable to the following attacks. Table \ref{tab:mia_vs_gia} presents a detailed comparison between Membership Inference Attack (MIA) and Gradient Inversion Attack (GIA), highlighting their goals, attack methods, targets, threat models, impacts, common defenses, severity levels, and use case in a hardware assurance setting.
\begin{itemize}
\item \textit{Membership Inference Attacks}: A membership inference attack is an attack technique that attempts to determine whether a specific data was part of the training set. This can leak sensitive information about the training data. For example, assuming there is a model trained on a specific SEM image, an attacker tries to apply that model to two node technologies (32nm and 90nm). If the model generates a very high confidence output for a particular technology node (e.g., 32nm), the attacker may conclude that the 32nm technology node is likely to have been used in the training set. In another example, the attacker may know the standard cell library of a foundry's process design kit (PDK) of a particular node. The attacker can use membership inference to identify that SEM images used during training were from that library. These attacks threaten the confidentiality of training data and can lead to the leakage of sensitive information. 
\makeExpprocessFig
\item \textit{Gradient Inversion Attacks}: Gradient inversion attacks refer to attempts to restore original data by exploiting updates shared during the FL process (e.g., model weights or gradients). For example, if an attacker (e.g., adversary server) has access to the gradient updates from other users during the FL process, they can reverse engineer them to restore the original data that contributed to generating those updates. Gradient Inversion Attacks can directly violate the confidentiality and privacy of data. Fig. \ref{fig:Gradient Inversion Attacks} shows the general concept of the gradient inversion attacks in the hardware assurance setting where SEM images are the input data. The adversary server initializes dummy data with random values and iteratively optimizes it to approximate the user’s private data. This optimization process minimizes the discrepancy between the dummy data's gradients and those of the user's actual data, effectively reconstructing sensitive information.
\end{itemize}

\section{Methodology}\label{sec:methodology}
\textcolor{black}{In this section, we discuss the applicability and potential benefits of FL and DL in hardware assurance. We have applied FL in IC reverse engineering with SEM. All experiments were conducted on an AMD EPYC ROME CPU with 32GB of RAM and an NVIDIA A100 GPU with 80GB of GPU RAM.}

\subsection{Dataset and Environment Setup}
\textcolor{black}{The segmentation task in FL requires access to SEM images of different layers of an IC, the design layout images, and the corresponding ground-truth images (i.e., ideal segmentation results) for training. The openly accessible REFICS dataset provides the necessary data to evaluate such a segmentation task for proof-of-concept~\cite{wilson2021refics}. The REFICS\footnote{Link: https://trust-hub.org/\#/data/refics} dataset consists of 800,000 synthetic SEM images taken at various Fields-of-View (FoVs) and dwelling times per pixel. These images were generated from the doping, metal, polysilicon, and contact layers of ICs from two technology nodes: 32nm and 90nm. For the purpose of our segmentation task, we utilized the doped layers of the 32nm technology node but we expect our results should generalize to other layers. From the tool packaged with the REFICS dataset, we generated a subset of 100 images with shot noise parameter set to 20 and 10 $\mu$sec/pixel dwelling time. The background and foreground means are 75 and 135, respectively, with a standard deviation of 20. To accommodate input requirements of our segmentation model, images were resized to 256$\times$256 pixels. 
}

\subsection{Workflow}
As shown in Fig. \ref{fig:exp_process}, our experimental workflow consists of three main processes designed to evaluate the effectiveness of FL in segmentation tasks and to demonstrate that FL alone is not sufficient to protect data privacy. First, data generation, where we prepare SEM image datasets divided among multiple clients to simulate a realistic FL environment. Second, the segmentation task using FL, in which clients collaboratively train a global segmentation model without sharing raw data. Third, gradient inversion attack, which is performed on the shared gradients to investigate the vulnerability of FL to privacy attacks and to assess whether sensitive data can be reconstructed from model updates. Through these steps, we aim to analyze not only the segmentation performance but also the privacy risks associated with FL in practical scenarios.

\subsubsection{Demonstrating Segmentation Task Performance
in FL}
In this study, we applied FedAVG \cite{mcmahan2017communication}, a fundamental federated learning algorithm, to train a segmentation model for IC reverse engineering using SEM images. In FedAVG, each client receives a global model from a central server and trains it locally using its own private data for a set number of epochs. After local training, each client sends updated parameters to the central server; the server then averages these results to update the global model.  This updated global model is then redistributed back to the clients, and this process is repeated in an iterative fashion.

To demonstrate the benefits of applying FL to segmentation tasks, we conducted several experiments. The goal is to compare segmentation performance between centralized learning (CL) using a single client and FL settings using multiple clients. FL settings allow multiple clients to collaborate on their own datasets, allowing for much larger datasets than CL. Through this, we aim to demonstrate the efficiency of FL and performance improvement through collaboration between multiple clients without sharing their sensitive data.

For this experiment, we prepared a dataset of SEM images from the 32nm technology node, ensuring an independent and identically distributed (iid) setup. In the CL training case, we set up a virtual Client 1 with a subset of the data, consisting of 10 SEM images referred to as ``Subset A''. We then trained a U-Net model on these data centrally, using only the virtual Client 1 and the ``Subset A" data without collaboration from other clients, representing the scenario where a single client with limited data trains its model independently. This case allows us to assess the baseline performance when data availability is limited.

In another CL training case, we set up a virtual Client 1 with a larger subset, consisting of 90 SEM images combining data from all other clients referred to as ``Subset K". We then trained a U-Net model on these data centrally, using only the virtual Client 1 and the ``Subset K" data. This case serves as a performance upper bound for comparison with FL settings, as centralized training with full data typically achieves the best possible performance at the cost of less privacy.

In the FL training case, we partitioned the dataset across nine virtual clients (Clients 1 to 9), each assigned 10 SEM images, referred to as ``Subsets A, B, C, D, E, F, G, H, and I", respectively. Each client trained a local model on its own subset, and model updates were aggregated at a central server following the FedAvg algorithm. This setup allowed a total of 90 images to be utilized collaboratively while maintaining data locality to preserve data privacy. All models are trained with hyperparameters found in Table \ref{tab:hyperparameters}.
\makeHyperparameter

\subsubsection{Exposing Privacy Vulnerabilities}
\makeExamplegiaFig
The goal of this study is to experimentally demonstrate that SEM images can be successfully reconstructed using only gradients shared in a FL environment using a gradient inversion attack (GIA). Through this, we would like to hypothesize that FL alone is not sufficient to protect privacy. In order to evaluate if FL environment is vulnerable to privacy attacks, we modified GIAs optimized for our first task, image segmentation, based on the existing techniques \cite{zhu2019deep} and \cite{geiping2020inverting}.

Most prior works on GIAs involve image classification tasks \cite{du2024sok}, whereas this study concerns image segmentation models. Since image segmentation models generate pixel-by-pixel image outputs, rather than simple class predictions like classification models, segmentation models are more structurally complex, and are thus relatively less vulnerable to GIAs. Therefore, in this study, we propose a GIA optimized for image segmentation tasks by modifying the loss function.

By combining the loss functions used in previous studies, we optimize our proposed GIA to be more suitable for image segmentation. The Mean Squared Error (MSE) loss function maximizes pixel-wise accuracy, enabling more precise reconstructions. On the other hand, the Cosine Similarity (CS) loss function is advantageous for capturing global structures and high-level features, helping to maintain the overall shape of the image. By combining these two loss functions, the proposed approach effectively reconstructs both fine patterns and the overall structure of SEM images. The trade-off between MSE and CS loss is controlled by the parameter $\alpha$, which can be interpreted as a regularization strength. A higher value of $\alpha$ places greater emphasis on pixel-wise accuracy, while a lower value prioritizes global structure preservation. Fig. \ref{fig:example_gia} shows an example of the evolution of our GIA on 32nm technology node data.
\begin{equation}
Loss = \alpha \times MSE + (1-\alpha) \times CS
\end{equation}
This allows the model to adapt to the specific needs of the task and data, and can be tuned during the training process to achieve the best results.

In this study, we assume that the server acts as an adversary in the FL, and perform a simulation to recover the original SEM image by collecting gradients or model weights transmitted from one of the clients. The adversary performs reverse engineering using the shared gradient from the client and evaluates the possibility of restoring sensitive original data with only the information delivered during the learning process of the model. As we applied the Federated Averaging (FedAVG) algorithm for the segmentation task, we aim to demonstrate how gradient inversion attacks can be carried out in a FedAVG setting, where only model weights, not gradients, are shared. To demonstrate the proof-of-concept of the attack, we assume users train the model with a single sample image and share the updated weights after completing one epoch of local training. This is based on findings from \cite{geiping2020inverting}, which demonstrated that although the GIA is possible with multiple images, the results are less accurate. Additionally, \cite{du2024sok} discussed the limitations of the GIA, emphasizing that it is predominantly effective with one image, a conclusion supported by other studies.

\makeSegresultss

Federated Stochastic Gradient Descent (FedSGD) and Federated Averaging (FedAVG) \cite{mcmahan2017communication} can both be used in federated learning, but they differ in how they approach gradient inverse attacks. In FedSGD, each client learns a model with private local data and then sends the gradients to the central server, which averages them to update the global model. In this process, the client directly transmits the gradient, so an attacker can use the gradient sent to the server to perform a gradient inversion attack that can infer the original data. On the other hand, in FedAVG, each client trains the model locally for several epochs, then sends the new model weights to the server, which averages the weights from each of the clients to update the global model. In this case, because FedAVG does not directly share the gradient but only transmits the weight, the attacker can use the difference between the new weight and the previous weight to estimate the gradient and perform a gradient inversion attack based on this.

To retrieve the gradients from these shared weights, we rely on the fact that the weight updates are a result of the gradient-based optimization algorithm, e.g., Stochastic Gradient Descent (SGD). By subtracting the previous model weights (Old Weights) from the updated model weights (New Weights), we can approximate the gradient using the formula: 
\begin{equation}
    \nabla L_i(\theta_i^t) \approx \frac{\text{Old Weights} - \text{New Weights}}{\eta}
\end{equation}
where $\eta$ is the learning rate. This allows us to infer the gradients, which are crucial for the GIA. Once we have the approximate gradients, we can then apply gradient inversion techniques to reconstruct the original SEM images.

\subsection{Evaluation}
To evaluate segmentation accuracy, we employ 
Intersection-over-Union (IoU) using the Jaccard index to measure shape localization accuracy, Mean Squared Error (MSE) to quantify pixel-wise differences from the ground truth, and Structural Similarity Index Measure (SSIM) to assess the structural fidelity of the reconstructed layout. Each listed metric focuses on a different facet of the image and, when aggregated, enables a holistic assessment of similarity to the ground truth image i.e., assess if most features are preserved. For evaluation, we used a separate hold-out set of 10 SEM images referred to as ``Subset J'' assigned to virtual Client 10. 
To evaluate GIA effectiveness, we compared the SEM image reconstructed from GIA to the original SEM image quantitatively using MSE, SSIM, and Peak Signal-to-Noise Ratio (PSNR).

\section{Results and Discussion}\label{sec:results}
In this section, we show results and discuss their implications. We analyze not only the segmentation performance but also the privacy risks associated with FL considering vulnerabilities to gradient inversion attacks (GIAs).

\makeExamplesegFig

\subsection{Segmentation Task Performance in FL}
\makeIoULossFigs

The MSE, IoU, and SSIM scores for each configuration are presented in Table \ref{tab:FLsegresult_combined}. The trained segmentation model in the 9-client FL scenario, which achieved scores of 0.955 IoU, 0.013 MSE, and 0.860 SSIM, demonstrated its ability to produce high-quality segmentation masks for unseen SEM images (shown in Fig. \ref{fig:example_seg}). Our results provide several key insights. As expected, CL training with limited data (10 images) produced the worst segmentation performance due to insufficient information for robust pattern recognition. This highlights the challenges of training accurate models when data availability is limited. FL with multiple clients significantly outperformed centralized training with 10 images, and performance improved as the number of clients increased, as evidenced by higher IoU and SSIM and lower MSE. This result underscores the benefit of collaborative training in FL, where distributed data contributions lead to enhanced model generalization while preserving client privacy. However, FL still underperformed compared to centralized training with full access to all 90 images, highlighting that, despite preserving privacy, FL may not fully match the performance achievable when all data is aggregated centrally.

In this study, we observed that FL achieved lower MSE than centralized training with 90 images. This occurs because MSE reflects pixel-wise similarity and may favor smoother, blurry predictions that avoid sharp pixel errors. While these predictions achieve lower MSE, they often fail to capture accurate object boundaries and structures, leading to reduced IoU and SSIM. In contrast, models from CL training with 90 images that produce sharper segmentation masks may incur higher MSE due to pixel-level deviations but better capture the shape and structure of the objects, resulting in higher IoU and SSIM. This highlights that MSE alone may not fully represent segmentation quality, and structural metrics like IoU and SSIM are essential to evaluate true segmentation performance.

For training time analysis, we assumed instantaneous communication (uploads and downloads) to focus on local training and server-side aggregation. Our measurements show that FL requires more training time than centralized training with 10 images, as it involves multiple rounds of local updates and global aggregation. Additionally, as the number of clients increases, the training time also increases due to the additional overhead of server-side model aggregation. However, FL was often faster than centralized training with 90 images, which requires more extensive computation on the entire dataset in each epoch. Furthermore, consistent with \cite{mehta2022federated}, when FL was trained with a total of 90 images, equivalent to CL training, both its performance and training time were worse than CL training, confirming the trade-offs between data privacy and efficiency.

Although FL allows model training without directly accessing clients' raw data, this privacy-preserving approach comes at the cost of higher computation and potentially lower performance compared to centralized learning on aggregated data. As \cite{mehta2022federated} mentioned, in some real-world scenarios, the computational cost of FL is not straightforward to compute. Thus, its use must be evaluated for each potential task.

Fig. \ref{fig:results_subfigures} presents the performance evaluation of a segmentation model in a FL setting, relative to a CL setup with a single client.
The first figure presents the ``Test Loss for Different Total Numbers of Clients.'' This plot demonstrates that the test loss converges to a lower value as the number of clients in the FL setup increases (from 1 to 9). These results indicate that as more clients participate in the training, the test loss decreases within the same total number of epochs, highlighting the advantage of the FL approach over the centralized learning (CL) setup with a single client.

The second figure illustrates the ``Test IoU for Different Total Numbers of Clients.'' This plot shows the IoU metric, a widely used evaluation criterion for segmentation tasks. Similar to the test loss, the IoU improves more rapidly as the number of clients in the FL setup increases, achieving higher values compared to the CL setup with a single client.

\makeGiametrics

Overall, these figures provide a comprehensive evaluation of the segmentation model’s performance and demonstrate the benefits of the FL approach in terms of better convergence to lower test loss and improved segmentation accuracy within the same number of training epochs, compared to the CL setup.

\makeExampleSSIMFig
\makeExampleMSEPSNRFig

Our experiments demonstrate that FL offers a promising privacy-preserving alternative to centralized learning for segmentation tasks, enabling multiple clients to collaboratively train models without exposing sensitive data. FL can significantly improve segmentation performance compared to single-client training on limited data, and that increasing the number of participating clients further enhances performance. However, FL still falls short of centralized training with full access to all data in terms of ultimate performance and training efficiency. These results highlight the practical trade-offs between maintaining data privacy and achieving optimal performance. These insights are essential for practitioners considering FL for segmentation tasks, particularly in fields where data privacy and distribution constraints are critical.

\subsection{Privacy Vulnerabilities}



As shown in Table \ref{tab:examplegiametrics} and Figures \ref{fig:example_SSIM} and \ref{fig:example_MSEPNSR}, the experimental results demonstrated that the GIA can successfully reconstruct SEM images in a FL environment, suggesting that FL alone is not sufficient to protect privacy. Our U-Net model is more complex than the ResNet model commonly used for image classification in most GIA studies. Despite these complexities posing additional challenges for the success of the GIA, we successfully reconstructed a SEM image. Furthermore, our image data has a higher resolution of 256x256 pixels, compared to the 224x224 pixel resolution, which is the highest used in previous studies where most employ even lower resolutions \cite{du2024sok}. This higher resolution, combined with the increased model complexity, poses additional challenges for the GIA. Nevertheless, our results indicate the success of the attack even under these more demanding conditions. Furthermore, while \cite{ziller2021differentially} were the first to demonstrate GIA on segmentation tasks, they did GIA in a setting that provided the attacker with the ground truth segmentation mask rather than a dummy mask (randomly initialized mask) during the GIA process to facilitate convergence and acceptable reconstructions for large models. In contrast, our scenario does not assume access to a ground truth segmentation mask reflecting a more realistic and severe attack. Despite this, our results \ref{fig:example_SSIM} and \ref{fig:example_MSEPNSR} show that the worst-case reconstruction is still recognizable to human observers. From a privacy perspective, this is critical due to the revealing of identifiable or sensitive features by partial or imperfect reconstruction can comprise a serious breach. A partially reconstructed image can reveal technology information such as technology node (e.g., 32nm vs 90nm), via sizes, and layering structures. Even if the reconstruction is not perfect at the pixel level, it still reveals sensitive content that could reveal functional aspects of the chip layout and design features. This finding emphasizes the severe privacy risks latent in FL-based segmentation operations, even in settings that are more challenging than those considered previously such as using lower complexity models and resolution images, and having ground truth segmentation masks. 

The presence of noise in the reconstructed image can negatively impact the SSIM score, even if the structural information is preserved. To mitigate the effect of noise and focus the SSIM evaluation on structural similarity, we applied the LASRE segmentation algorithm \cite{wilson2020lasre}, which is both unsupervised and robust to the presence of noise, to the reconstructed image prior to calculating the SSIM. LASRE was also chosen considering the fact that the adversary may not have access to ground-truth data to train state-of-the-art segmentation algorithms. This step helps remove the noise-induced variations in pixel values while preserving the structural information in the image. By applying this algorithm, we can obtain a more accurate assessment of the structural similarity between the original mask and reconstructed images, without the confounding effects of noise.

To ensure a consistent and meaningful interpretation of the Mean Squared Error (MSE) metric, we first normalized the pixel values of the original and reconstructed images to the range of 0 to 1. This normalization step is crucial, as the raw pixel values can vary depending on the image format and bit depth, which would affect the absolute scale of the MSE. By constraining the pixel values to the [0, 1] range, we were able to calculate the MSE on a standardized scale, allowing for a more direct comparison of the reconstruction quality across different images. Once the MSE was computed on the normalized pixel values, we then derived the Peak Signal-to-Noise Ratio (PSNR) based on this MSE score.

\section{Future Research Directions}\label{sec:Discussion and Future Research Directions}




There are a number of potential research opportunities for future work. Blue team opportunities include exploring the benefits and limitations of other privacy-preserving techniques that may be combined with FL, such as Differential Privacy (DP), Generative Adversarial Networks (GAN), and combinations. Frameworks for integrating such techniques and combinations in hardware security and metrics to help identify and address weaknesses could help researchers and practitioners increase defenses against AI attacks. 

Red team opportunities include exploration of gradient inversion attacks for other AI models used in hardware assurance, such as transformer, language, or multimodal models, and applying GIA on multiple images to better understand the attack’s practicality in more realistic scenarios. In addition, there are a variety of other attacks to be explored, such as model stealing, membership/property inference attacks, model poisoning, and combinations. In model stealing, adversaries could effectively reverse-engineer trained models and produce counterfeits at low-to-no development costs. In membership/property inference attacks, sensitive IP from the training data could be leaked. In model poisoning, adversaries could degrade model integrity and reduce user trust. In addition, we also plan to apply our attack to other chip layers and evaluate our attack's robustness in non-independent and identically distributed (non-iid) settings.

In any case, there is a great need for comprehensive frameworks or taxonomies of AI attacks, defenses, and countermeasures specifically in the context of hardware security and assurance. There is also a great need for publicly-accessible resources in this area, including case studies, datasets, tools, and reproducibility/comparison studies. Such future directions are crucial for facilitating the development of novel methodologies and increasing collaboration among researchers and practitioners in academia, industry, and government.

\section{Conclusions}\label{sec:conclusions}

This study explored the potential and limitations of applying deep learning and federated learning (FL) to hardware assurance. While FL has been widely used in fields like healthcare, IoT, and finance, its application in hardware assurance remains limited due to the challenges discussed in Section \ref{sec:background}. This highlights the need for further research and specialized approaches in this area. Our segmentation task using SEM images demonstrates that FL allows multiple participants to collaboratively train models without sharing raw data, improving model performance. However, it does not fully eliminate privacy risks. As shown in our Gradient Inversion Attack experiment, sensitive data can still be reconstructed. These challenges indicate that FL alone is insufficient to guarantee privacy, emphasizing the need for additional privacy-preserving techniques.

\footnotesize{
\bibliographystyle{IEEEtran}
\bibliography{Sections/references}
}

\end{document}